\documentclass[onecolumn]{revtex4}
\usepackage{graphicx}

\begin{document}

%%%%%%%%%%%%%%%%%% title page information %%%%%%%%%%%%%%%%%%
\title{Phase-coherent repetition rate multiplication of a mode-locked laser from 40 MHz to 1 GHz by injection locking}

\author{D. Kielpinski}

\address{Centre for Quantum Dynamics, Griffith University, Nathan QLD 4111, Australia}

\author{O. Gat}

\address{Racah Institute of Physics, Hebrew University, Jerusalem IL-91904, Israel}

\email{dave.kielpinski@gmail.com} %% email address is required



\begin{abstract}
We have used injection locking to multiply the repetition rate of a passively mode-locked femtosecond fiber laser from 40 MHz to 1 GHz while preserving optical phase coherence between the master laser and the slave output. The system is implemented almost completely in fiber and incorporates gain and passive saturable absorption. The slave repetition rate is set to a rational harmonic of the master repetition rate, inducing pulse formation at the least common multiple of the master and slave repetition rates.
\end{abstract}

\maketitle

%%%%%%%%%%%%%%%%%%%%%%% References %%%%%%%%%%%%%%%%%%%%%%%%%
%\begin{thebibliography}{99}
%\bibitem{gallo99} K. Gallo and G. Assanto, ``All-optical diode based on second-harmonic generation in an asymmetric waveguide,'' \josab {\bf 16,} 267--269 (1999).
%\end{thebibliography}

%%%%%%%%%%%%%%%%%%%%%%%%%%  body  %%%%%%%%%%%%%%%%%%%%%%%%%%
\section{Introduction}

Many applications of mode-locked laser sources require spectral resolution of the individual frequency comb lines produced by the laser, including optical frequency counting \cite{Bartels-Diddams-10GHz-comb}, calibration of astronomical spectrographs \cite{Benedick-Kartner-visible-astro-comb}, broadband spectroscopy \cite{Thorpe-Ye-broadband-cavity-ringdown}, and line-by-line optical waveform generation \cite{Cundiff-Weiner-optical-AWG-rev}. The most common mode-locked lasers have repetition rates in the tens of MHz, so that the comb lines are not easily resolved by, e.g., diffraction from a grating. To overcome this problem, many investigators are now striving to produce mode-locked lasers with repetition rates in the GHz range while maintaining the excellent frequency stability needed for these applications.

In principle, high repetition rate is readily obtained by reducing the length of the laser resonator. This approach has been demonstrated for solid-state \cite{Bartels-Diddams-10GHz-comb} and fiber lasers \cite{McFerran-Newbury-Er-passive-ML-2GHz}, but is inherently limited by technical constraints on miniaturization and the nonlinear dynamics of the mode-locking mechanism. These difficulties become more prominent as the repetition rate is pushed higher. Alternatively, one can spectrally filter the frequency comb from a low-repetition-rate mode-locked laser using a Fabry-Perot cavity with large free spectral range (FSR) \cite{Benedick-Kartner-visible-astro-comb, Steinmetz-Udem-FP-rep-rate-filter}. The Fabry-Perot length is stabilized to an exact multiple of the laser resonator length so that the frequency comb lines separated by a multiple of the FSR are transmitted. The average power output is reduced by the same factor as the repetition rate is increased, lowering signal-to-noise in the applications discussed above. While such sources are known to exhibit low long-term frequency drift, their optical phase noise and time-domain properties are yet to be thoroughly investigated.

Harmonically mode-locked lasers and their variants offer tantalizingly high repetition rates, but exhibit frequency noise that is intrinsically worse than the solutions discussed above. They also exhibit undesirable pulse-to-pulse energy variations known as supermode noise, which can degrade the stability of supercontinuum generation used for, e.g., absolute frequency stabilization. The incorporation of a Fabry-Perot etalon within the laser cavity can reduce these drawbacks \cite{Harvey-Mollenauer-etalon-active-harmonic-ML, Gee-Delfyett-etalon-active-harmonic-ML}, but the resulting pulses are at least 5 ps long, making them unsuitable for the wideband frequency-comb applications discussed above.

Here we demonstrate phase-coherent repetition-rate multiplication of a mode-locked master laser by injection-locking a mode-locked slave laser at a rational harmonic of the master laser repetition rate. The system is implemented entirely in fiber, except for the delay line, and uses only off-the-shelf optical components. An output repetition rate over 1 GHz is obtained, 25 times the master laser repetition rate of 40.2 MHz. The frequency comb generated by the slave is found to be optically coherent with the master laser comb, as well as having highly stable repetition rate and low supermode noise. The output pulses are suitable for coherent supercontinuum generation after amplification and soliton compression. The principle demonstrated here may be extended to yet higher repetition rate multiplication and shorter pulse duration, offering a straightforward route to high-repetition-rate optical frequency combs.

\section{Principle of operation}

In the situation described here, a passively mode-locked master laser injection-locks a passively mode-locked slave laser at a rational harmonic of the master laser repetition rate $\nu_M$. The output of the slave is then a train of identical mode-locked pulses at a multiple of $\nu_M$, and the output pulses are optically phase-coherent with the master laser. As shown experimentally below, the repetition rate can be multiplied by a factor of at least 25, from a typical fiber laser repetition rate of 40 MHz to an output repetition rate over 1 GHz.

The master laser spectrum consists of sharp spikes at frequencies $\nu =  n_M \nu_M + \nu_\mathrm{M0}$, where $\nu_\mathrm{M0}$ is the master offset frequency and $n_M$ is an integer. The allowed slave modes occur at frequencies $\nu = n_S \nu_S + \nu_\mathrm{S0}$, where $\nu_S$ is the free spectral range of the slave resonator, $\nu_\mathrm{M0}$ is the slave resonator offset frequency, and $n_S$ is an integer. The slave resonator length is tuned so that
\begin{equation}
\nu_S/\nu_M = p/q, \quad \mbox{ $p, q$ integers, $p, q$ relatively prime} \label{vernier}
\end{equation}
Eq. (\ref{vernier}) is referred to here as the "vernier condition" and ensures that $\nu_S$ and $\nu_M$ are related by a rational harmonic. Hence master laser comb lines spaced by $p \nu_M$ can injection-lock the slave laser when the resonance condition
\begin{equation}
\nu_\mathrm{S0} = \nu_\mathrm{M0} + \frac{m p}{q} \nu_M  \quad \mbox{$m$ integer} \label{carrier}
\end{equation}
is satisfied. The output from the slave resonator then consists of identical mode-locked pulses at repetition rate $\nu_\mathrm{out} = p \nu_M$.

Our experimental system tolerates significant deviations from the ideal injection-locking scenario, in which conditions \ref{vernier} and \ref{carrier} are exactly satisfied. Gain competition in the slave resonator causes frequency pulling of the slave resonator modes to the injected frequencies, an effect that strengthens with increasing injection power. The pulling effect relaxes the injection-locking conditions, leading to improved robustness of the system as compared to, e.g., passive Fabry-P\'erot repetition-rate multipliers. The mismatch tolerance has been theoretically evaluated by Margalit and co-workers for the case $p/q = 1$, i.e., when the master and slave repetition rates are approximately equal \cite{Margalit-Haus-injection-lock-MLL-theory}. For typical soliton fiber lasers, they calculated that the allowable repetition rate mismatch was a few kHz, while the offset frequency mismatch $|\nu_\mathrm{S0} - \nu_\mathrm{M0}|$ could be several hundred kHz. Our experimental system exhibits mismatch tolerances on this order, although the locking range is not systematically studied here.

A related repetition-rate multiplier, which was not configured to retain optical phase coherence, was previously studied by Margalit and co-workers \cite{Margalit-Eisenstein-40GHz-injected-harmonic-ML}. High-repetition-rate pulses at $\nu_M \sim 1$ GHz were injected into a slave laser with much lower free spectral range $\sim 10$ MHz. Output pulses at repetition rates up to $10 \nu_M$ could be obtained by varying the relative repetition rates of master and slave. It was believed that rational-harmonic injection-locking was responsible for this behavior. In contrast to our work, only pulses of $\sim 6$ ps duration were obtained, so that coherent supercontinuum generation would not have been possible \cite{Dudley-Coen-fiber-SC-rev}. Moreover, the very narrow frequency spacing of the slave modes allowed injection locking to take place on various sets of slave modes as the master offset frequency fluctuated, effectively changing the value of $m$ in Eq. (\ref{carrier}). The relative offset frequency between master and slave was not measured or stabilized, so master-slave optical phase coherence could easily be lost as the injected modes hopped from one set of slave modes to the next. The present experiments maintain injection-locking on a fixed set of slave modes and are shown to maintain optical phase coherence. However, we were also able to observe behavior similar to that of \cite{Margalit-Eisenstein-40GHz-injected-harmonic-ML} by deliberately detuning $\nu_S$ by a few parts in $10^3$ from the vernier condition (\ref{vernier}).

\section{Experimental realization}

\begin{figure}
\centerline{\includegraphics[width=10cm]{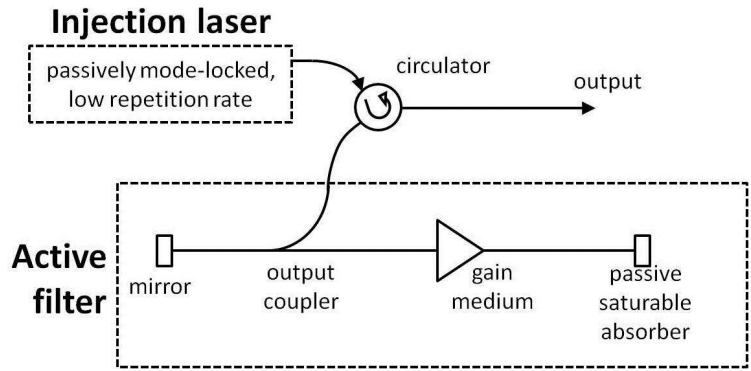}}
\caption{Schematic of the experimental setup.}
\label{schem}
\end{figure}

Our experimental repetition-rate multiplication system is shown in Figure \ref{schem}, and used an Er-doped master fiber laser at 1550 nm to injection-lock an Er-doped slave fiber laser. The slave laser incorporated a semiconductor saturable absorber mirror (SESAM) and was designed for low round-trip group-velocity dispersion. Seed light from the master laser was injected through a circulator into the slave resonator output coupler. With the exception of a single free-space delay arm, the entire system was constructed in fiber from standard commercially available components. When operated without injection light and at low pump power, the slave laser exhibited standard soliton mode-locking dynamics with an optical spectrum spanning $\sim 7$ nm full-width at half-maximum (FWHM). Harmonic self-mode-locking was obtained when the pump power to the slave exceeded the threshold for pulse breakup, as commonly occurs in SESAM mode-locked lasers. However, in the absence of injection light, harmonic mode-locking was only obtained up to repetition rates $\sim 400$ MHz and with supermode noise far larger than obtained with injection.

The master laser was a commercial passively mode-locked Er-doped fiber laser (Precision Photonics) emitting $\sim 150$ fs pulses at a repetition rate of $\nu_M = 40.2$ MHz. We have previously used this laser to generate coherent octave-spanning supercontinuum, showing that it is adequate for the optical frequency comb applications discussed above \cite{Kielpinski-Feder-octave-SC-pulse-generation}. The master laser was amplified in a home-built chirped-pulse Er-doped fiber amplifier to a maximum power of $\sim 25$ mW. After dechirping, the amplified seed light passed through a circulator and was injected into the slave resonator via a fused-fiber output coupler with 20\% coupling ratio. The seed pulse bandwidth was much larger than the bandwidth available in the slave resonator. The chirp of the seed light was carefully adjusted to maximize the bandwidth of the output pulses from the system.

The slave laser was built from standard single-mode fiber (SMF-28 and HI1060) and was designed to have small anomalous group-delay dispersion. It incorporated $\sim 20$ cm of Er-doped fiber (Liekki Er110-4/125) with 110 dB/m attenuation at the 976 nm pump wavelength, pumped through a fiber wavelength-division multiplexer with up to 400 mW from a 976 nm fiber-coupled laser diode (Lumics LU0980M400). One end of the slave laser was butt-coupled to a semiconductor saturable absorber mirror (SESAM) with 18\% saturable and 10\% nonsaturable absorption and 2 ps recovery time (BATOP SAM-1550-30-x-2ps). At the other end, the fiber was cleaved and the light was collimated and retroreflected into the resonator by a gold mirror on a translation stage. This arrangement served as a free-space delay line, allowing fine adjustment of the slave repetition rate, and was the only portion of the entire system not implemented in fiber. The overall laser free spectral range amounted to $\sim 80$ MHz. The optical output of the system was emitted from the same coupler port used for injection; the circulator at this port served to separate the filter output from the source light. The other coupler output, containing most of the injected light as well as output light, passed through a standard isolator to prevent back-reflection into the slave resonator.

Stable injection locking could only be achieved if the offset frequency mismatch is actively stabilized. This result is roughly in accord with the estimated offset frequency mismatch tolerance of \cite{Margalit-Haus-injection-lock-MLL-theory}. To stabilize the offset frequency, we employed a variant of the RF locking technique that is commonly used for stabilizing a laser to an external resonator \cite{Black-rf-lock-tutorial}. The injection light was frequency-modulated at $\nu_\mathrm{RF} \sim 2$ MHz by an in-fiber electro-optic modulator. The injection light rejected by the slave laser coupler was detected by a photodiode and the resulting photocurrent was demodulated at $\nu_\mathrm{RF}$, yielding a low-frequency error signal approximately proportional to the offset frequency mismatch. The error signal was fed back through a servo loop of $\sim 1$ kHz bandwidth to a piezoelectric actuator that controlled the master laser resonator length. Stable locking was readily achieved despite the presence of unwanted slave output light at the photodiode.

\section{System performance}

We obtained stable injection-locked operation with output repetition rates exceeding 1 GHz, a multiplication factor of 25 relative to the master laser. At 1 GHz output repetition rate, the system output power was 12.5 mW, or 35 times higher than the power expected from passive filtering of the seed light (see below). The pulse duration of 470 fs was near the transform limit imposed by the observed 5.5 nm spectral bandwidth. Observation of a high-quality heterodyne signal between the master laser and the system output confirmed that optical phase coherence was preserved by the repetition rate multiplication process. The output repetition rate was found to have high spectral purity.

Injection locking could only be achieved above a critical seed power, as predicted for harmonic injection locking by the thermodynamic theory of Weill and co-workers \cite{Weill-Fischer-injected-MLL}. In the absence of injection locking, the slave laser could not be mode-locked at the operating current. Here we present a complete characterization of the system configuration at 1 GHz output repetition rate, but similar performance was found for a wide variety of multiplication factors ranging from 3$\times$ to 25$\times$, indicating the flexibility of our system. We plan to investigate the injection-locking threshold dynamics in future work.

\begin{figure}
\centerline{\includegraphics[width=\columnwidth]{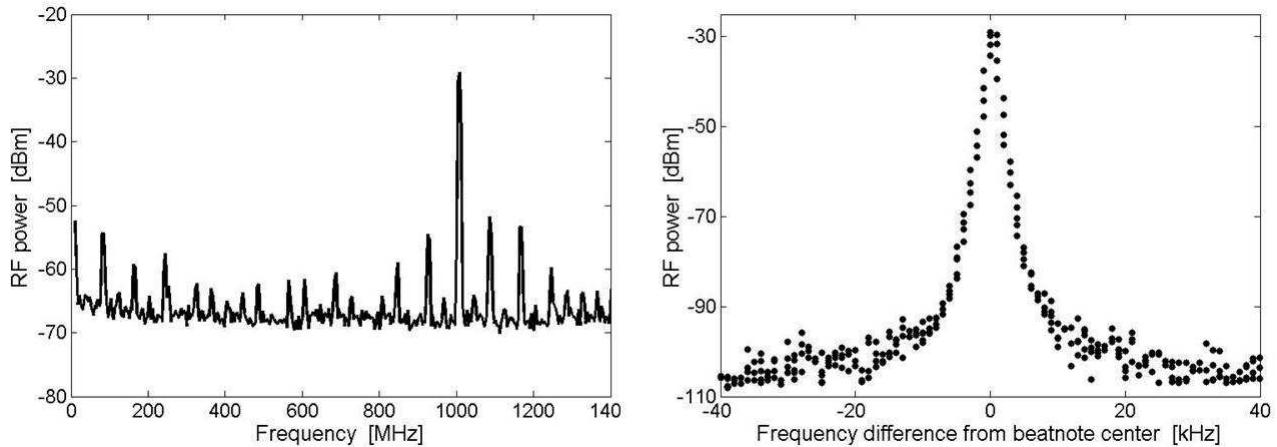}}
\caption{Left: RF spectrum of the system output. The main peak at 1.01 GHz is associated with the pulse repetition rate. Supermode noise peaks occur at the fundamental repetition rates of master and slave resonators, but are suppressed by $>25$ dB. Right: Measurement of the 1.01 GHz peak at higher frequency resolution. The -3 dB width of the signal is consistent with the 1 kHz resolution limit of the RF spectrum analyser.}
\label{rfpic}
\end{figure}

The RF spectrum of the system output (Figure \ref{rfpic}, left) was obtained by photodetection of the system output. The repetition rate of the system is indicated by the strong peak at 1.01 GHz. Supermode peaks with amplitudes at least $25$ dB lower than the main peak were observed at multiples of the master laser repetition rate. The supermode peaks presumably arise from the effect of residual injection light on the laser dynamics. At lower multiplication factors the supermode noise tends to be more strongly suppressed; for instance, at output repetition rate of 442 MHz (11$\times \nu_M$) the suppression is over $33$ dB. Observation of the 1 GHz peak at higher frequency resolution (Figure \ref{rfpic}, right) demonstrates the high spectral purity of the repetition rate. The -3 dB width of the signal is consistent with the 1 kHz resolution limit of the RF spectrum analyser, indicating that the repetition rate multiplication adds less than 1 kHz additional frequency noise to the repetition rate.

\begin{figure}
\centerline{\includegraphics[width=\columnwidth]{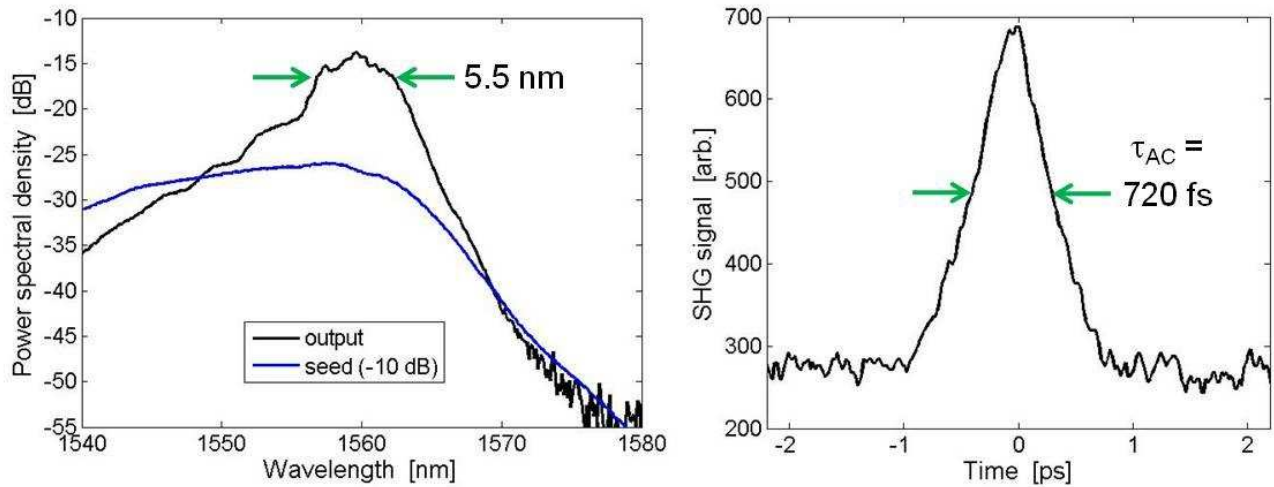}}
\caption{Left: Optical spectrum of the injection-locked output (black line) and the seed light (blue line). The traces are offset for clarity. The FWHM bandwidth is $\sim 5.5$ nm. Right: Autocorrelation trace of the amplified output from the active filter. }
\label{optcorr}
\end{figure}

The optical spectrum of the injection-locked system is shown in Figure \ref{optcorr} (left). The optical bandwidth was found to be 5.5 nm FWHM, consistent with the bandwidth of the slave laser in the absence of injection locking, but much smaller than the seed bandwidth of several tens of nm. Proper matching of chirp between the seed pulses and the circulating slave pulses was essential for obtaining this bandwidth. The pulse duration of the system output was estimated from the second-harmonic autocorrelation trace shown in Figure \ref{optcorr} (right). The autocorrelation duration was found to be 720 fs FWHM. Assuming a sech$^2$ envelope, we obtain a pulse duration of 470 fs, at the transform limit of the optical spectrum. We calculate that the circulating pulses in the slave laser have soliton number of approximately 1.

\begin{figure}
\centerline{\includegraphics[width=8cm]{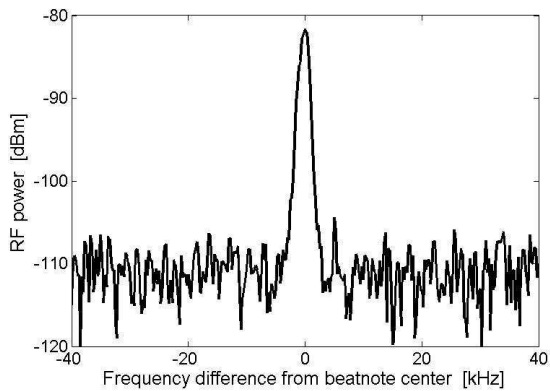}}
\caption{Measurement of the optical frequency noise added by repetition rate multiplication. An optical beatnote is obtained by heterodyning the system output with the seed oscillator. The -3 dB width of the beatnote is consistent with the 1 kHz resolution limit of the RF spectrum analyzer. The beatnote strength is sufficient to demonstrate optical phase coherence in the multiplication process.}
\label{optbeatnote}
\end{figure}

The optical phase coherence between the master laser and the system output was demonstrated by a heterodyne measurement. The system output was phase-modulated at a modulation frequency $\nu_H \sim 2$ MHz using an electro-optic phase modulator and was interferometrically combined with the master laser light on a fast photodiode. After synchronization of the master and output pulse trains, the RF spectrum of the photocurrent exhibited beatnotes at $\nu_M \pm \nu_H$. The beatnotes disappeared if either the master or output paths was blocked, showing that they arose from optical beating between the master and output. The spectrum of one of these heterodyne beatnotes is shown in Fig. \ref{optbeatnote}. The -3 dB width of the beatnote is consistent with the 1 kHz resolution limit of the RF spectrum analyzer, indicating that $<1$ kHz noise is added to the master laser offset frequency $\nu_\mathrm{M0}$ by the multiplication process. Moreover, the beatnote signal-to-noise ratio is over 25 dB, showing that optical phase locking without cycle slips over an essentially infinite duration can be achieved by a feedback loop of bandwidth 100 kHz \cite{Telle-optical-frequency-measurement-noise}, more than adequate for operation of a self-referenced fiber frequency comb \cite{Newbury-Swann-fiber-frequency-comb-rev}. For practical purposes, we can conclude that optical phase coherence is maintained by the repetition rate multiplication.

\section{Conclusion}

We have demonstrated repetition-rate multiplication of a mode-locked laser to 1 GHz, a factor of 25, by injection locking at a rational harmonic. The multiplication process is optically phase coherent and adds minimal frequency noise to the repetition rate. The output pulse duration appears to be controlled by the slave laser bandwidth and is similar to that of other SESAM mode-locked fiber lasers. The system is implemented almost completely in fiber, using standard telecom components, making it robust and durable. The upper limit of the multiplication factor is unknown and is to be studied in future work.

The repetition-rate multiplication technique appears applicable to any mode-locked laser technology, including solid-state and diode lasers as well as fiber lasers. Replacing the SESAM by a faster saturable absorption mechanism, such as nonlinear polarization rotation, would increase the slave bandwidth and presumably allow for multiplication of shorter pulses. This improvement would aid use of the multiplication technique in self-referenced optical frequency combs, allowing for novel comb applications. For instance, the frequency of an unknown laser source could be determined and locked coarsely with respect to the multiplied comb, then locked with higher precision to the seed comb, combining tight locking with large servo capture range.

\begin{figure}
\centerline{\includegraphics[width=10cm]{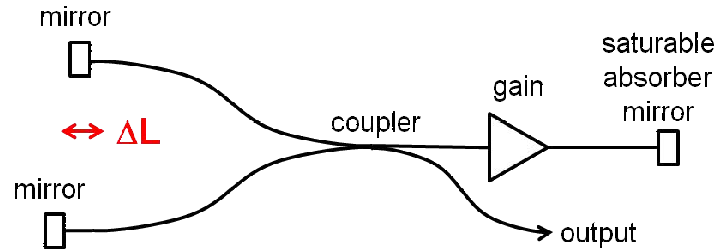}}
\caption{Schematic of \textsf{Y}-cavity laser exploiting self-injection to achieve high repetition rate. The vernier condition is satisfied for correct adjustment of the relative path length $\Delta L$ between the two cavity branches at left.}
\label{twobranch}
\end{figure}

The principle of rational-harmonic injection locking could be used to design novel passively mode-locked lasers that operate at high repetition rate. In the \textsf{Y}-shaped cavity design of Fig. \ref{twobranch}, each of the two branches on the left combine with the common branch on the right to form two coupled mode-locked lasers with lengths $L$ and $L + \Delta L$, where $\Delta L \ll L$. When the two coupled laser cavities satisfy the vernier condition (\ref{vernier}), each laser cavity will injection-lock the other and the entire system will oscillate at repetition rate $c/(2 \Delta L)$, much larger than the repetition rate $c/(2 L)$ of the individual cavities. Because the carrier frequency of the system is free to change in response to changing cavity lengths, the injection process will be less subject to environmental phase noise and might operate robustly even without active stabilization of the branch lengths.

\section{Acknowledgments}

This work was funded by the Australian Research Council (DP0773354), by the US Air Force Office of Scientific Research (FA2386-09-1-4015), and by the Israeli Science Foundation (1002/07).

\end{document}